\documentclass[twocolumn]{aa}

\usepackage{graphicx}
\usepackage{txfonts}
\usepackage{natbib}
\bibpunct{(}{)}{;}{a}{}{,}
\newcommand{\lastcor}{}
\begin{document}
   \title{Curvelet analysis of asteroseismic data}
   \subtitle {I: Method description and application to simulated sun-like stars}

   \author{	P. Lambert\inst{1,2},
			S. Pires\inst{3,1},
			J. Ballot\inst{2,4},
			R. A. Garc\'{\i}a\inst{1,2},
			J.-L. Starck\inst{3,1}
			\and
			S. Turck-Chi\`{e}ze\inst{1,2}     
			}

   \offprints{P. Lambert}

   \institute{	DSM/DAPNIA/Service d'Astrophysique,
              		CEA/Saclay,
              		91191 Gif-sur-Yvette Cedex,
              		France \\
              		\email{lambertp@cea.fr} 
        \and
				AIM -- Unit\'{e} Mixte de Recherche CEA - CNRS - Universit\'{e} Paris VII --
UMR n\degr 7158,
				CEA/Saclay,
            		91191 Gif-sur-Yvette Cedex, 
				France 
        \and
             	 	DSM/DAPNIA/Service d'Electronique des D\'{e}tecteurs et
d'Informatique,
              		CEA/Saclay,
              		91191 Gif-sur-Yvette Cedex,
              		France 
	   \and
				Max-Planck-Institut f\"ur Astrophysik,
				Karl-Schwarzschild-Str. 1,
				Postfach 1317,
				85741 Garching, Germany        
	}

		\date{Received 2005 November 18; Accepted 2006 March 31}

		\abstract
			{
The detection and identification of oscillation modes (in terms of their $\ell$,
$m$ and successive $n$) is a great challenge for present and future
asteroseismic space missions. The ``peak tagging" is an important step in the
analysis of these data to provide estimations of stellar oscillation mode
parameters, i.e., frequencies, rotation rates, and further studies on the
stellar structure.
			}
			{
To increase the signal-to-noise ratio of the asteroseismic spectra computed from
time series representative of MOST and CoRoT observations (30- and 150-day
observations).
			}
			{
			 We apply the curvelet transform -- a recent image processing technique which
looks for curved patterns -- to echelle diagrams built using asteroseismic power
spectra. In this diagram the eigenfrequencies appear as smooth continuous
ridges. To test the method we use Monte Carlo simulations of several sun-like
stars with different combinations of rotation rates, rotation-axis inclination
and signal-to-noise ratios. 
			}
			{
			The filtered diagrams enhance the contrast between the ridges of the modes
and the background allowing a better tagging of the modes and a better
extraction of some stellar parameters. Monte Carlo simulations have also shown
that the region where modes can be detected is enlarged at lower and higher
frequencies compared to the raw spectra. Even more, the extraction of the mean
rotational splitting from modes at low frequency can be done \lastcor{more
easily} than using the raw spectrum.   
			}
              	{} 
		
		\keywords{Stars: oscillations -- Methods: data analysis -- Techniques: image
processing}
  	
	\authorrunning{P. Lambert et al.}
 	\maketitle

\section{Introduction}

Helioseismology -- the study of solar oscillations -- is a powerful probe of the
structure and dynamics of the Sun which has provided great improvements in our
understanding of stellar evolution and structure \citep[][ and references
therein]{TurckChieze1993, JCD2002}. Those successes push the community to apply
seismic techniques to other stars, opening the doors to asteroseismology, the
study of stellar oscillations. These oscillations have already been observed
from ground and space. The ground-based observations are limited by the
day-night cycle, which introduces aliases in the observations, but allow to use
Doppler velocity measurements. They have provided data with \lastcor{sufficient
quality} to detect solar-like oscillations \citep[see][ and references
therein]{BouchyCarrier2003, BeddingKjeldsen2003}. To reduce the aliases,
multi-site campaigns have been carried out but they are too short to have a good
frequency resolution. Space photometry missions and ground-based velocity
networks must be used to provide observations of stellar oscillations without
these limitations. With the current MOST\footnote{Microvariability and
Oscillations of STars \citep{Matthews1998}} and WIRE\footnote{Wide-field Infra
Red Explorer \citep{Buzasi2000}} satellites and the future
COROT\footnote{Convection Rotation and planetary Transits \citep{Baglin2001}}
mission asteroseismology is blooming. However, we still have to deal -- in the
case of solar-like oscillations -- with very small signal-to-noise ratio
(hereafter $S/N$) observations as a consequence of the weakness of the
luminosity variations. Moreover, stars cannot be spatially resolved yet. Only
global oscillation modes can be observed. In addition, we cannot have access  to
the rotation rates and the rotation-axis inclination separately. Without knowing
these two key stellar properties, the tagging of the modes in terms of their
properties ($\ell, m$) and successive $n$ may be extremely difficult. In fact,
the main problem to face will not be to fit the peaks (``peak-bagging'') but to
provide a good description of the model to be fitted after having put the
correct labels on the modes (``peak tagging''). To do this, it has been proposed
to use the echelle diagram where the modes follow ridges depending on the
stellar properties. To improve the S/N ratio \citet{Bedding2004} proposed to
filter this diagram by a vertical \lastcor{smoothing}. However the
\lastcor{smoothing} works well only when the ridges are quasi-vertical which
means a very good \textit{a priori} knowledge of the large difference and is
restricted to the asymptotic part of the spectrum. We propose here to follow a
similar approach but using new mathematical denoising techniques better suited
to the study of curved ridges.

At the end of the last decade, the application of mathematical transforms based
on wavelets to analyze astronomical images has been widely developed. The first
wavelet algorithms were well adapted to treat images with isotropic elements.
However, this description presented a limitation in the context of astrophysics,
where objects such as filaments or spiral structures exhibit a highly
anisotropic character (in shape and scale). New transforms, the ridgelet
\citep{Candes1998} and curvelet transforms \citep{CandesDonoho1999, Starck2002},
were then developed to deal efficiently with such objects. Astrophysical
applications (image denoising) of this technique have been presented in
\citet{Starck2003, Starck2004} to analyze images of gravitational arcs, the
Saturn rings or the CMB (Cosmic Microwave Background) map.

In this paper we suggest to use the curvelet transform to analyze asteroseismic
observations (more precisely the stellar echelle diagrams), in order to improve
the ``peak tagging'' of the oscillation modes and even the resultant ``peak
bagging''. To illustrate the application of this denoising technique in the
asteroseismic case, we have performed Monte Carlo simulations of ideal
asteroseismic data contaminated by different levels of stochastic noise. We
start in Sect.~2 by a quick reminder of the properties of stellar oscillation
modes in the solar-like case and the construction of the echelle diagram. In
Sect.~3 we introduce multiscale transforms, in particular the ridgelet and the
curvelet transforms. In Sect.~4, the simulated data of a star with an
oscillation spectrum similar to the Sun but with different rotation axis
inclinations and rotation rates, are presented. In Sect.~5 we discuss the
results obtained in the simulations.

\section{Properties of solar-like oscillations}

\begin{figure}
	\includegraphics[scale=0.35,angle=0]{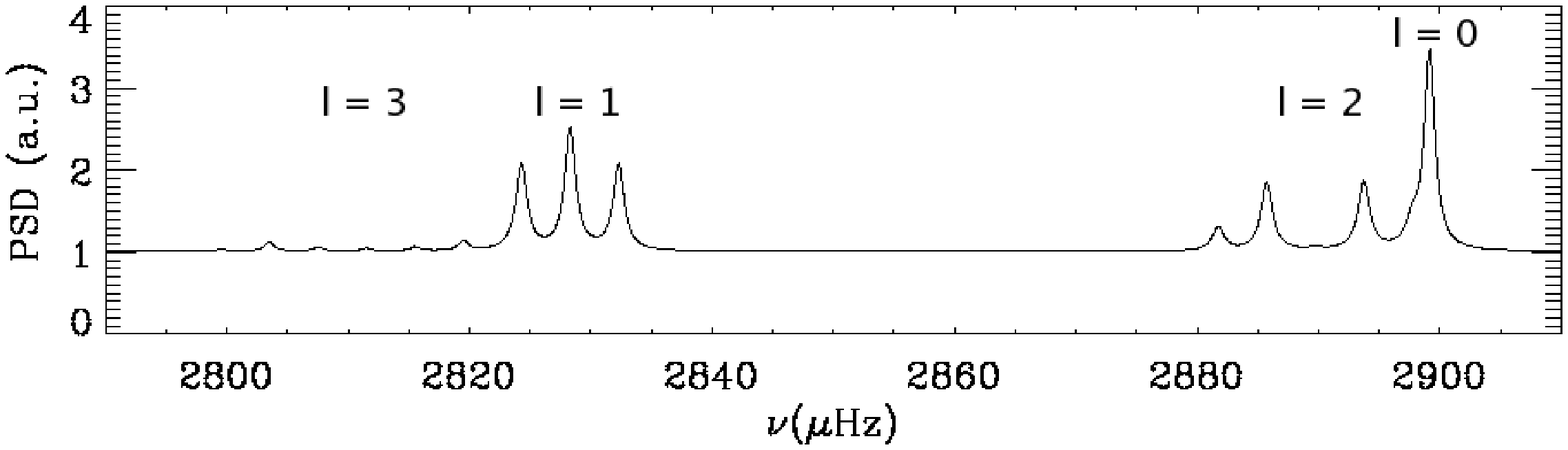}
	\includegraphics[scale=0.35,angle=90]{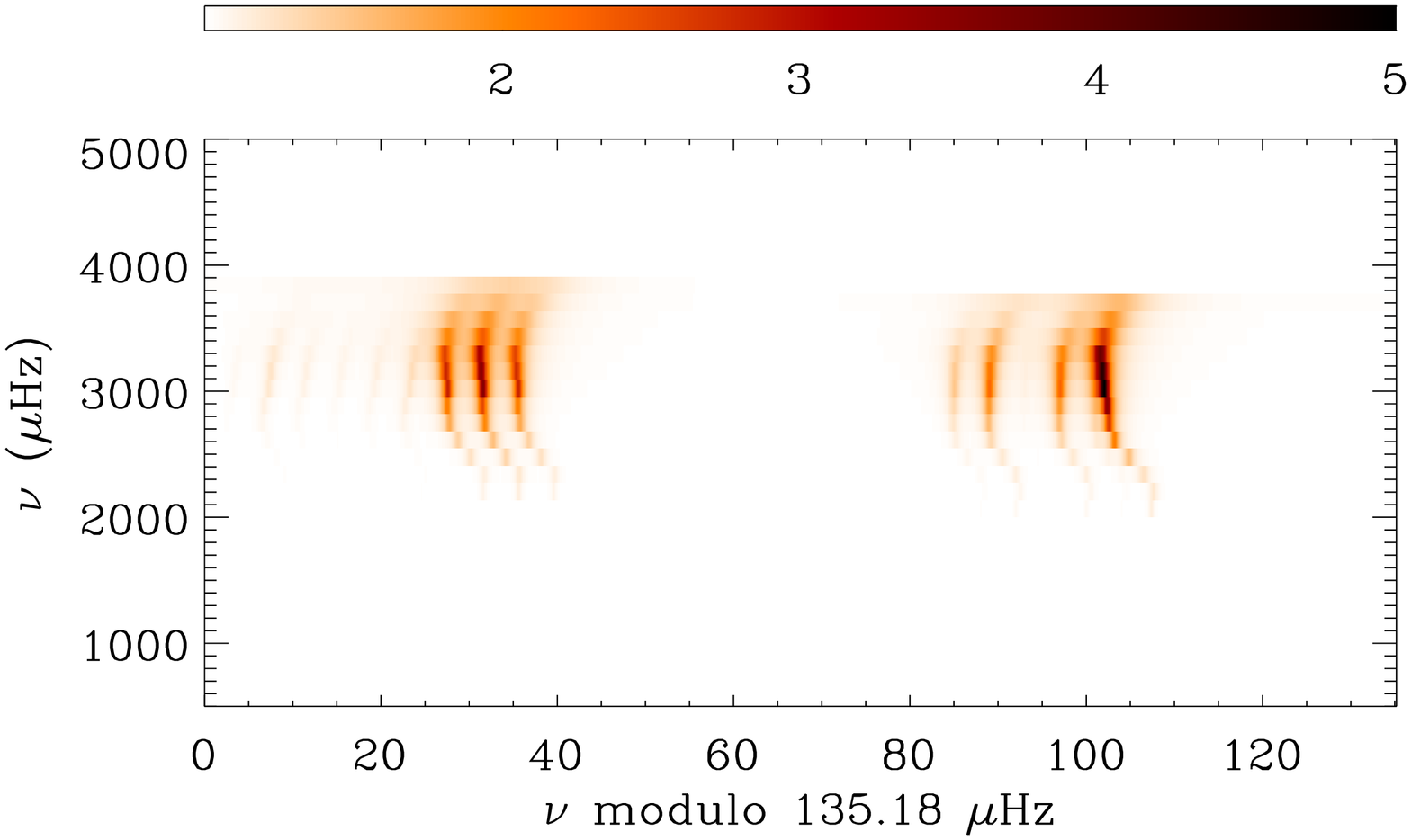}
	\caption{Portion of the theoretical spectrum (top) and echelle diagram (bottom)
for a sun spinning ten times faster than the Sun and seen under an angle of
50$\degr$. This is the ideal power spectrum used in the simulations described in
Sect.~5.}
	\label{theorique}
\end{figure}

Only low-degree stellar oscillation modes can be detected and observed with the
present generation of instruments. The asymptotic theory of oscillation modes
($n\gg \ell$) is then adequate and can be used to study them. First order
\citep{Tassoul1980} and second order developments \citep{Vorontsov1991,
Lopes1994, Roxburgh2000a, Roxburgh2000b} have been made to describe solar and
stellar oscillations. In the case of solar-like stars, where p-modes
predominate, the frequencies can be developed as:
\begin{equation}\label{secondordre}
	\nu_{n,\ell} \approx \Delta\nu_0 \big(  n+\frac{\ell}{2}+\frac{1}{4}
+\alphaup(\nu) \big) + \frac{\Delta\nu_0}{4\pi^2\nu_{n,\ell}}\big((\ell +
1/2)^2A + \psi\big) 
\end{equation}
in this expression $\ell$ and $n$ are respectively the degree and the radial
order of the modes and  
\begin{eqnarray*}
\tau_c			&=&	\int_{r_{in}}^{r_{out}}\frac{dr}{c_s} \\
\Delta\nu_0	&=&	\frac{1}{2\tau_c} \\
A				&=&	\frac{1}{4\pi^2\nu_{n,\ell}}\big(\frac{c_s(R_\star)}{R_\star} -
\int_{r_{in}}^{r_{out}}\frac{dc_s}{dr}\frac{dr}{r}\big) 
\end{eqnarray*}
$c_s$ is the internal stellar sound speed, $\alphaup$ is a phase-shift term and
$\psi$  is a function which allows to take into account the gravitational
potential in the central region \citep{Lopes1994}. From the asymptotic approach,
we can extract general properties of modes and better understand the physics
hidden in the frequencies behavior. The large frequency spacing, defined as
$\Delta\nu_{n,\ell}=\nu_{n+1,\ell} - \nu_{n,\ell}$, tends asymptotically to
$\Delta\nu_0$, related to the mass and radius of the star; the small frequency
spacing, $\delta_{\ell,\ell+2}\nu=\nu_{n,\ell}-\nu_{n-1,\ell+2}$, can be
approximated to first order by
$(4\ell+6)\Delta\nu_0/(4\pi^2\nu_{n,\ell})\int_0^{R_\star}\frac{dc_s}{dr}\frac{dr}{r}$.
This variable is related to the derivative of the sound speed and enhances the
effect coming from the central regions, providing constraints on the age of the
star. Finally the second difference is defined as
$\delta_2\nu=\nu_{n+1,\ell}-2\nu_{n,\ell}+\nu_{n-1,\ell}$. Its variations
provide information about the extent of the convective zone \citep{Monteiro2000,
Ballot2004} or the helium abundance in the stellar envelope \citep{Basu2004}.

Under the rotation effects the azimuthal order $m$ ($-\ell \leqslant m \leqslant
\ell$) is needed to characterize the oscillation spectrum. If the angular
velocity $\Omega$ is uniform \citep{Ledoux1951}, the mode frequencies are
asymptotically approximated by:
\begin{equation}\label{unifangvel}
	\nu_{n,\ell,m}\approx\nu_{n,\ell}+m\Omega/2\pi = \nu_{n,\ell}+m\delta\nu
\end{equation}
where $\delta\nu$ is the rotational splitting. Equation~\ref{unifangvel} shows
that modes are ($2\ell+1$)-times degenerated among the azimuthal order: a single
peak in the spectrum becomes a multiplet. Its corresponding structure depends on
the rotation rate, the inclination axis of the star and its stochastic
excitation. The solar-like mode lifetimes (a few days) are expected to be much
shorter than the length of the future space observations (a few months). In
consequence, the relative amplitude ratios inside a multiplet will only depend,
in average, on the inclination angle and the spacing between these different
m-components \citep{GizonSolanki2003}. Thus if the different m-components of a
multiplet can be identified and tagged with the correct $(\ell,m)$, they can
provide a good estimation of both the rotation-axis inclination $i$ and the
rotational splitting $\delta\nu$, allowing a better mode parameter extraction
through the fitting of the spectra. The effect of the stochastic excitation on
an isolated mode could be minimized by computing the average of these parameters
on several modes \citep[see for example the
n-collapsogramme;][]{BallotYale2004}.
 
Equation~\ref{secondordre} shows that the even ($\ell=0,2$) and odd ($\ell=1,3$)
modes have respectively almost the same frequency, only separated by the small
spacing $\delta_{\ell,\ell+2}\nu$. In addition, they are separated regularly in
frequency by the large spacing $\Delta \nu_{n,\ell}$. This property allows us to
build the so-called echelle diagram \citep{Grec1983}, which is currently used to
identify modes for solar-like oscillations. It is a 2D representation of the
power spectrum where this one is folded onto itself in units of the large
spacing. In such representation the modes appear as almost locally vertical
ridges (see Fig.~\ref{theorique}). The echelle diagram is a powerful tool for
the ``peak tagging'' since assigning the correct $(\ell,m)$ values to the peaks
is easier when the multiplet structure is well identified in this diagram. The
successive $n$ values are obtained from each individual horizontal line. 

    \section{Multiscale Transforms}
        \subsection{The Wavelet Transform}

\begin{figure}
    \centering
    \includegraphics[scale=0.245]{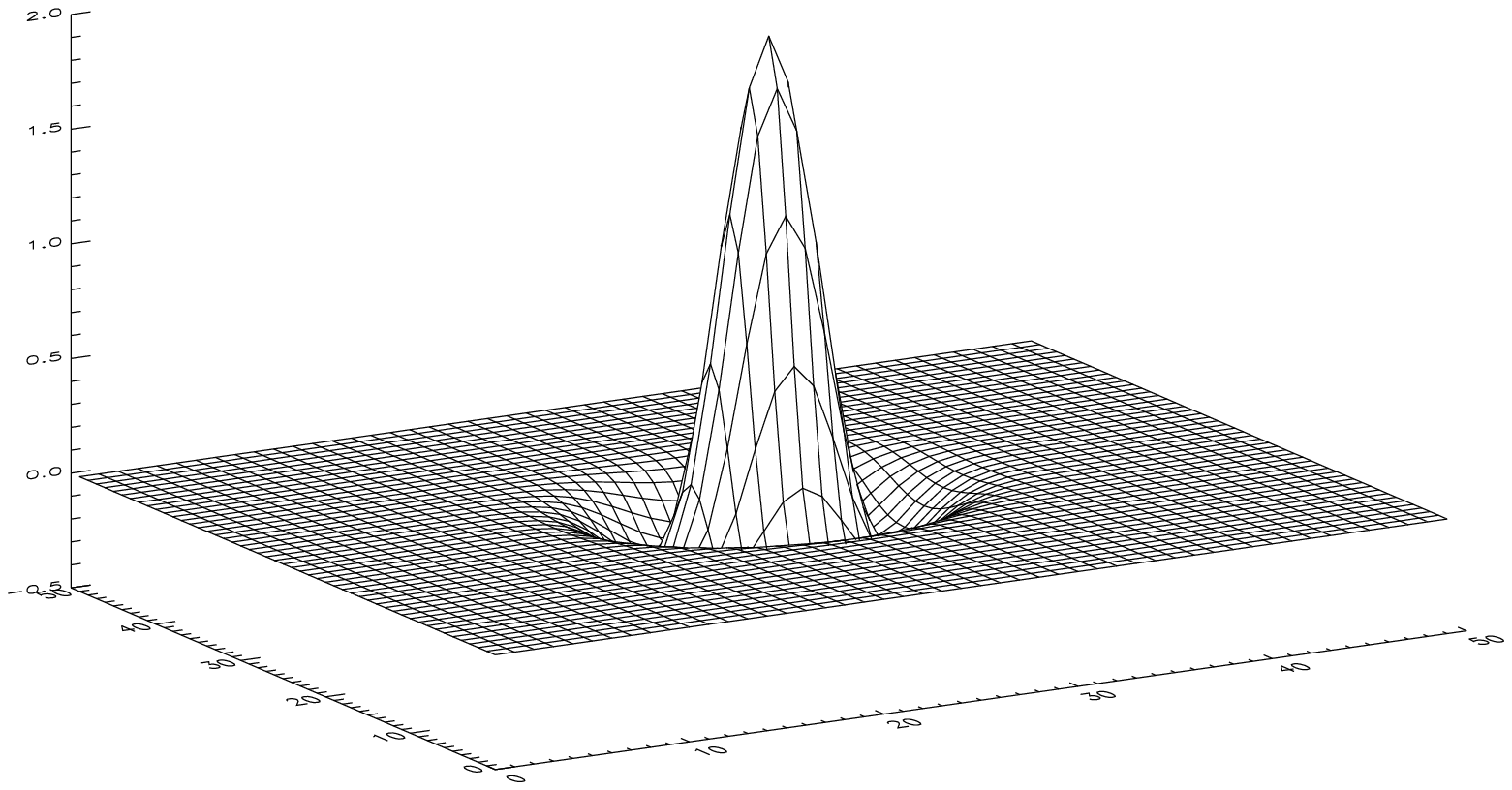}
    \includegraphics[scale=0.245]{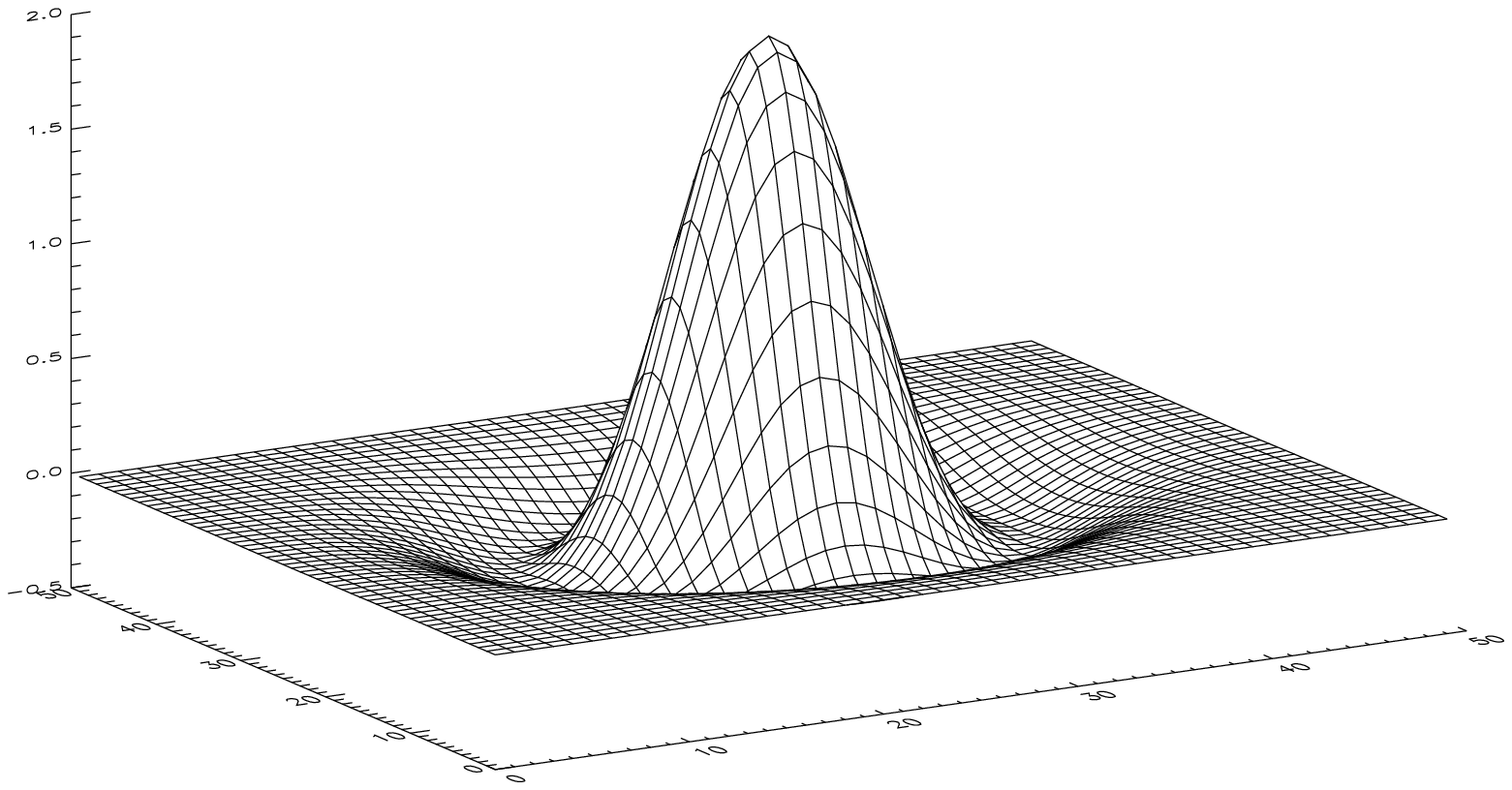}
    \includegraphics[scale=0.245]{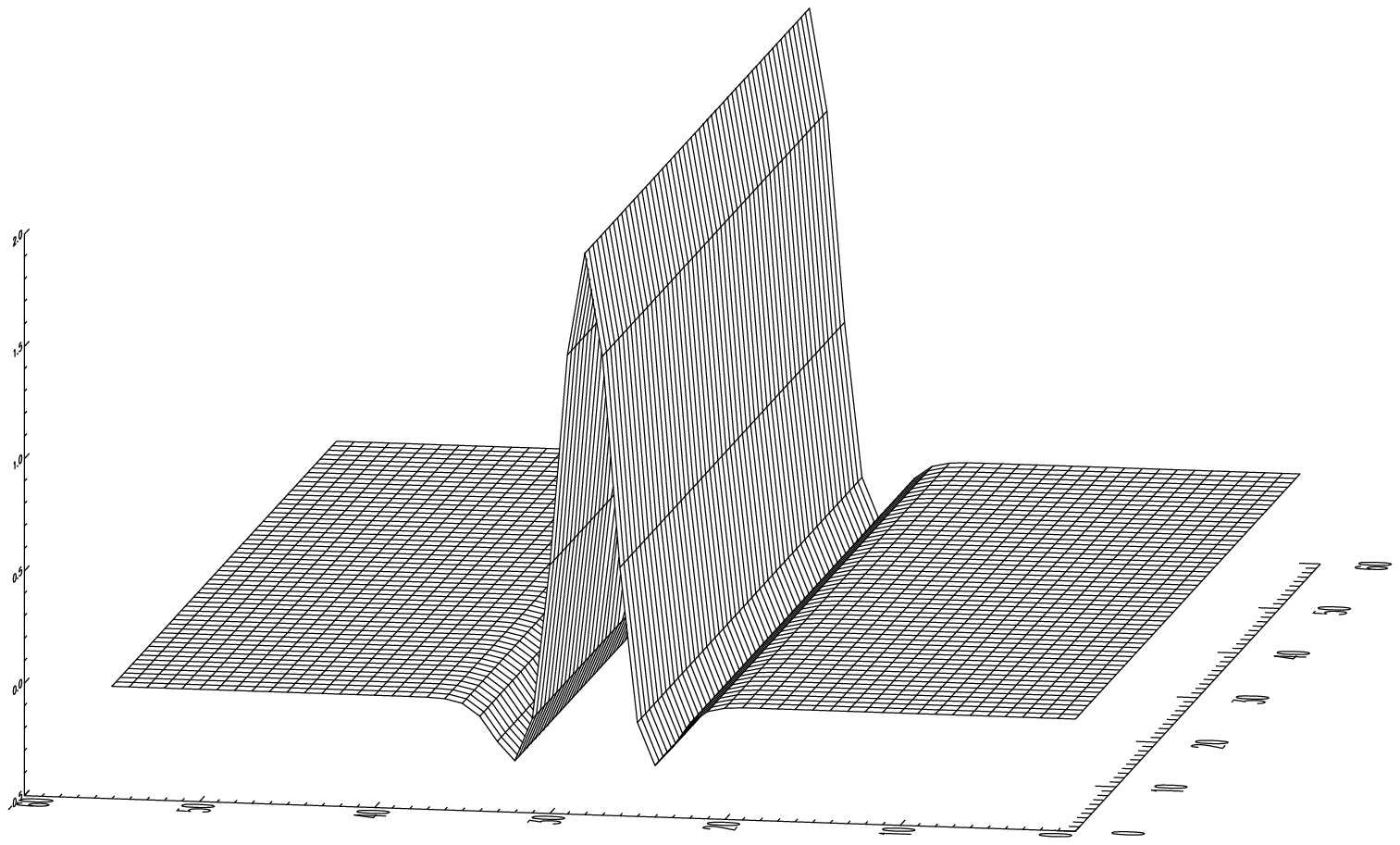}
    \includegraphics[scale=0.245]{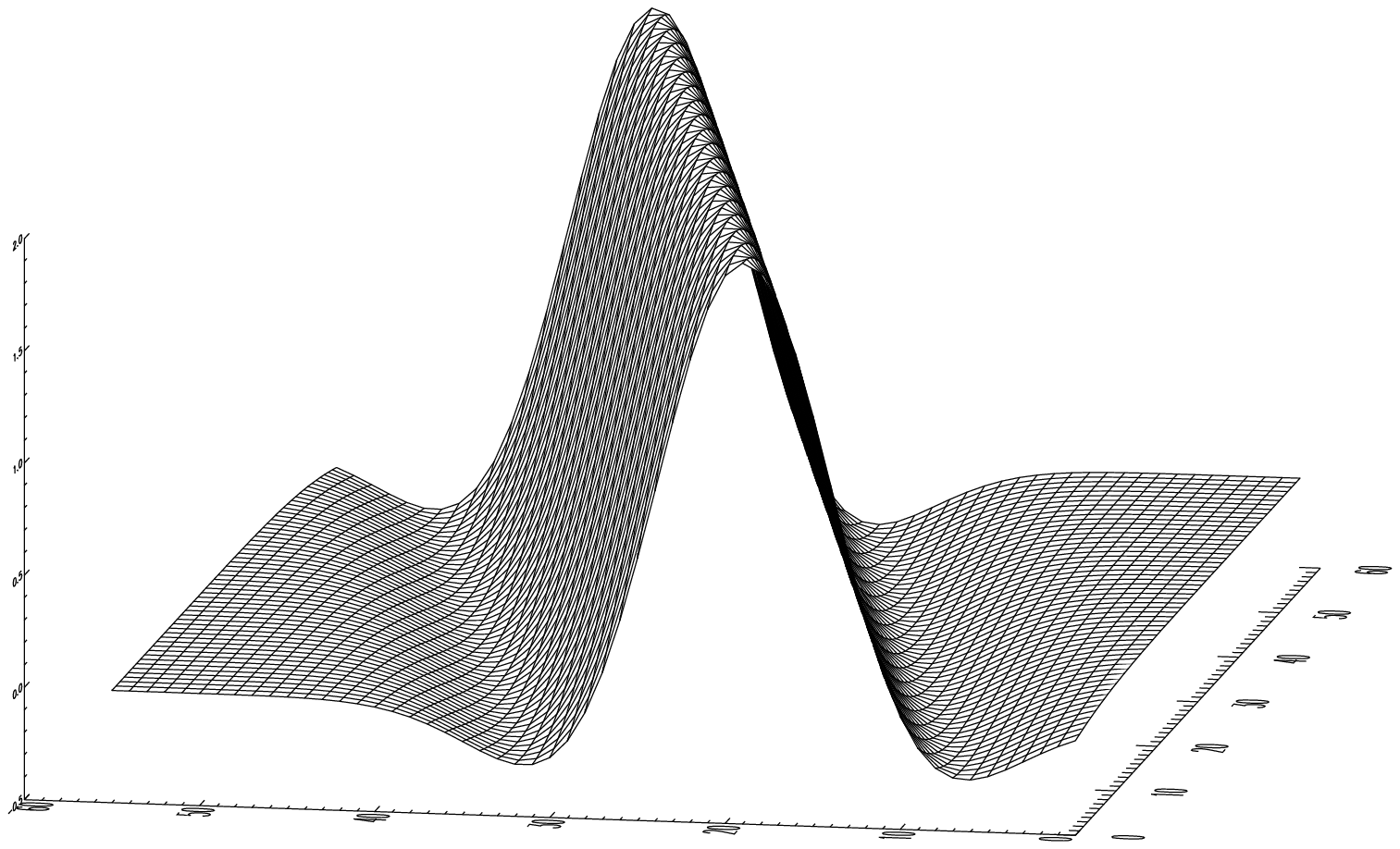}
    \caption{Examples of 2D wavelets (top panels) and ridgelets (bottom panels).
The top right wavelet has a greater scale parameter than this on the left. The
bottom right ridgelet has different orientation and width than the left one.}
    \label{examples}
\end{figure}

The wavelet transform provides a framework for decomposing images into their
elementary constituents across scales by reducing the number of significant
coefficients necessary to represent an image. The continuous wavelet transform
of a 2D signal is defined as:

\begin{eqnarray}
   W(a,b_i, b_j) = \frac{1}{\sqrt{a}}\int\!\!\!\int
f(x,y)\psi^*\left(\frac{x-b_i}{a},\frac{y-b_j}{a}\right)dxdy
\end{eqnarray}
where $W(a,b)$ are the wavelet coefficients of the function $f(x)$, $\psi(x)^*$
is the conjugate of the analyzing wavelet, $a>0$ is the scale parameter and $b$
is the position parameter. The continuous wavelet transform is the sum over all
the positions of the signal $f(x,y)$ multiplied by the scaled and shifted
versions of the wavelet $\psi((x-b_i) / a,(y-b_j) / a)$ (cf.
Fig.~\ref{examples}, top panels). This process produces wavelet coefficients
that are a function of scale and position.

However, the classical wavelet transform only address a portion of the whole
range of interesting phenomena: isotropic features at all scales and locations.
One of the drawbacks of the two-dimensional wavelet transform is that it does
not achieve an efficient analysis of images which present high anisotropy. For
instance, the wavelet transform does not efficiently approximate 2D edges, since
a large number of large wavelet coefficients, scale after scale, are required,
making difficult its analysis. In order to solve this problem two new
mathematical transforms, namely the ridgelet transform and the curvelet
transform, were introduced.

        \subsection{The Ridgelet transform}

The ridgelet transform was developed to process images including ridges elements
\citep{Candes1998}. It provides a representation of perfectly straight edges.
Given a function $f(x_1,x_2)$, the representation of this latter is the
superposition of elements of the form
$a^{-1/2}\psi((x_1\cos\theta+x_2\sin\theta-b)/a)$, where $\psi$ is a wavelet,
$a>0$ a scale parameter, $b$ a location parameter and $\theta$ an orientation
parameter. The ridgelet is constant along lines
$x_1\cos\theta+x_2\sin\theta=\mathrm{const}$, and transverse to these ridges it
is a wavelet. Thus, contrary to a unique wavelet transform, the ridgelet has two
supplementary characteristics: a length, equal to this of the image and an
orientation, allowing the analysis of an image in every direction and so
exhibiting the edge structure. Fig.~\ref{examples} (bottom panels) shows two
examples of ridgelets. The problem is that in the nature edges are typically
curved rather than straight so ridgelets alone cannot yield an efficient
representation.

	\subsection{The Curvelet transform}
		\subsubsection{Description}

\begin{figure}
    \centering
    \includegraphics[scale=0.32]{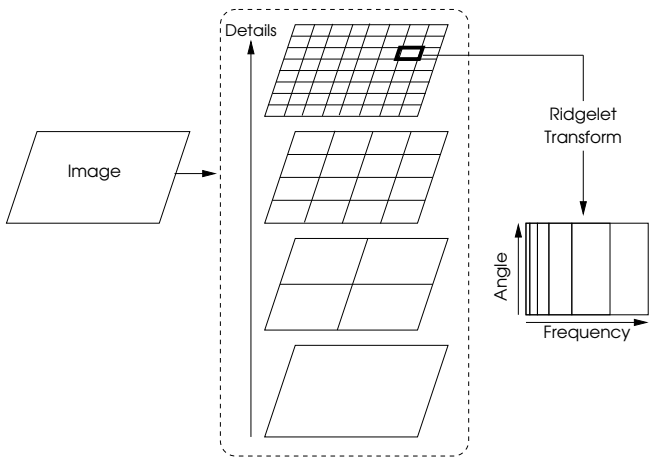}                        
    \caption{Sketch illustrating the curvelet transform applied to an image. The
image is decomposed into subbands followed by a spatial partitioning of each
subband. The ridgelet transform is applied to each block. The finest details
correspond to the highest frequencies.}
    \label{curveletgraphe}
\end{figure}

Ridgelets can be adapted to represent objects with curved edges using an
appropriate multiscale localization: at a sufficiently fine scale a curved edges
can be considered as almost straight. \citet{CandesDonoho1999} developed the
curvelet transform using ridgelets in this localized manner.
Fig.~\ref{curveletgraphe} shows the different steps of the curvelet analysis of
an image:

\begin{enumerate}
	\item Image decomposition into subbands: as a set of wavelets bands through a
2D isotropic wavelet transform. Each band corresponds to a different scale.
	\item Smooth partitioning: each subband is partitioned into squares -- blocks
--, whose size is appropriate to each scale. The finest is the scale, the
smaller are the blocks.
	\item Ridgelet analysis: it's applied to each square.
\end{enumerate}

The implementation of the curvelet transform offers an exact reconstruction and
a low computational complexity. Like ridgelets, curvelets occur at all scales,
locations and orientations. Moreover contrary to ridgelets, which have a given
length (the image size) and a variable width, the curvelets have also a variable
length (the block size) and consequently a variable anisotropy. The finest the
scale is, the more sensitive to the curvature the analysis is. As a consequence,
curved singularities can be well approximated with very few coefficients.

		\subsubsection{Denoising images: filtering curvelet coefficients}

To remove noise a simple thresholding of the curvelet coefficients has been
applied to select only significant coefficients. One possible thresholding of a
noisy image consists in setting to $0$ all non-significant curvelet coefficients
$\tilde c_{i,j,l}$, $i$, $j$ and $l$ respectively the indexes of the line, row
and scale: it is the so-called hard-thresholding:
\begin{eqnarray}
	\tilde c_{i,j,l} = \left\{ \begin{array}{ll} 	\mbox{1} & \mbox{if }  	c_{i,j,l}
\mbox{ is significant} \\ 
											\mbox{0} & \mbox{if } 	c_{i,j,l} \mbox{ is not significant}
\end{array} \right.
\end{eqnarray}
Commonly, $c_{i,j,l}$ is significant if the probability that the curvelet
coefficient is due to noise is small, i.e., if the curvelet coefficient is
greater than a given threshold. A basic problem remains: the choice of the
threshold. Usually, this threshold is taken equal to $k\sigma_j$, where
$\sigma_j$ is the noise standard deviation at the scale $j$ and $k$ is a
constant taken equal to 5 in our filterings.

Simple thresholding of the curvelet coefficients is very competitive
\citep{Starck2002} with ``state of the art'' techniques based on wavelets,
including thresholding of decimated or undecimated wavelet transforms.

\section{Simulation of data}

\begin{figure*}
    \centering
    \includegraphics[scale=0.65]{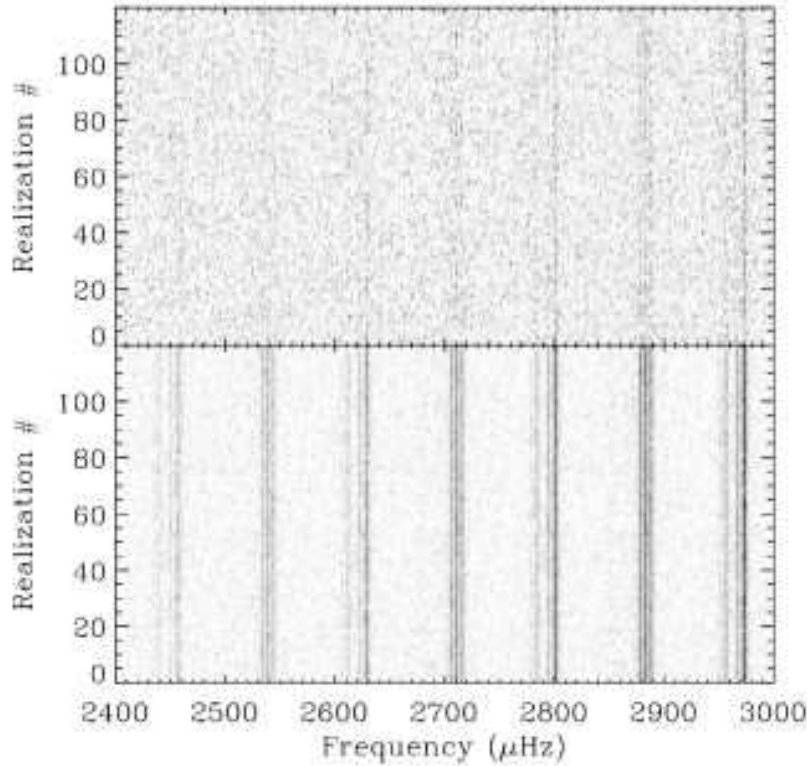}
    \caption{Effect of the curvelet denoising on the mode visibility for
$S/N=5$. Each picture shows 120 realizations out of the 500 done in our Monte
Carlo simulation. Each horizontal line corresponds to a single realization. The
top panel is the raw spectra and the bottom is the curvelet filtered one. 
\label{montecarlo}}
\end{figure*}

To characterize the curvelet denoising technique applied to the asteroseismic
data, we have simulated typical solar-like observations varying different
parameters: S/N ratios, observational lengths, rotation-axis inclinations,
rotation rates... With this approach we know the input parameters in advance and
we can evaluate the quality of the results given by the curvelet analysis and
its limits.

In the simulations shown in this paper, we use the oscillation spectrum of a
star similar to the Sun but seen under different conditions. Different rotation-axis inclinations ($i=50\degr$ and $90\degr$) and rotation
rates ($\Omega= \Omega_{\sun}$, $5\Omega_{\sun}$, and $10\Omega_{\sun}$) have
been considered. An ideal power spectrum were constructed first. Only the modes
$\ell\le3$, $n=12$--$25$ were simulated. The mode parameters -- frequencies
($\nu$), amplitudes ($A$) and widths ($\Gamma$) -- were obtained from the
analysis of GOLF (Global Oscillations at Low Frequency) data \citep{Garcia2004}.
The amplitudes were corrected to take into account the difference between
intensity and velocity observations. Modes were simulated with symmetrical
Lorentzian profiles as the asymmetry is expected to be at the level of the
noise. Following the method described in \citet{FierryFraillon1998}, a
multiplicative noise, a $\chi^2$ with 2 d.o.f. statistics, has been introduced
to reproduce the stochastic excitation of such modes \citep[see
also][]{Anderson1990}. The $S/N$ ratio of the ``resultant'' raw power spectrum
was defined as the maximum of the bell-shaped p-mode power (i.e. the highest
simulated p mode) divided by the noise dispersion. The simulated background is
flat assuming that it has been previously fitted and removed as it is usually
done for the Sun \citep{Harvey1985}.

Several Monte Carlo simulations have been performed for each ideal spectrum.
Realistic $S/N$, with values ranging from 5 to 15, have been used to cover a
wide range of situations (compatible with what it is expected, \citep[see
][]{Baglin2001}). In each realization of the Monte Carlo simulation the same
level of noise has been randomly added to the corresponding ideal spectra.
Therefore all the realizations, in a given Monte Carlo simulation, have the same
$S/N$ ratio. The simulated spectra have been computed for two resolutions,
$\approx0.38$ and $\approx0.077~\mu$Hz, corresponding respectively to 30-day and
150-day observations. The first are representative of MOST observations and the
short CoRoT runs while the latter are of the same length than the long CoRoT
runs.

Simulations of other stars, like some potential main CoRoT targets, with
different masses, ages and, in consequence, internal structures have been made.
The results have already been presented and discussed during the CoROT workshops
\#8 and \#9 obtaining the same qualitative results. For the sake of clarity,
they are not shown here.

\section{Discussion}
\begin{figure*}
    	\includegraphics[scale=0.35,angle=90.]{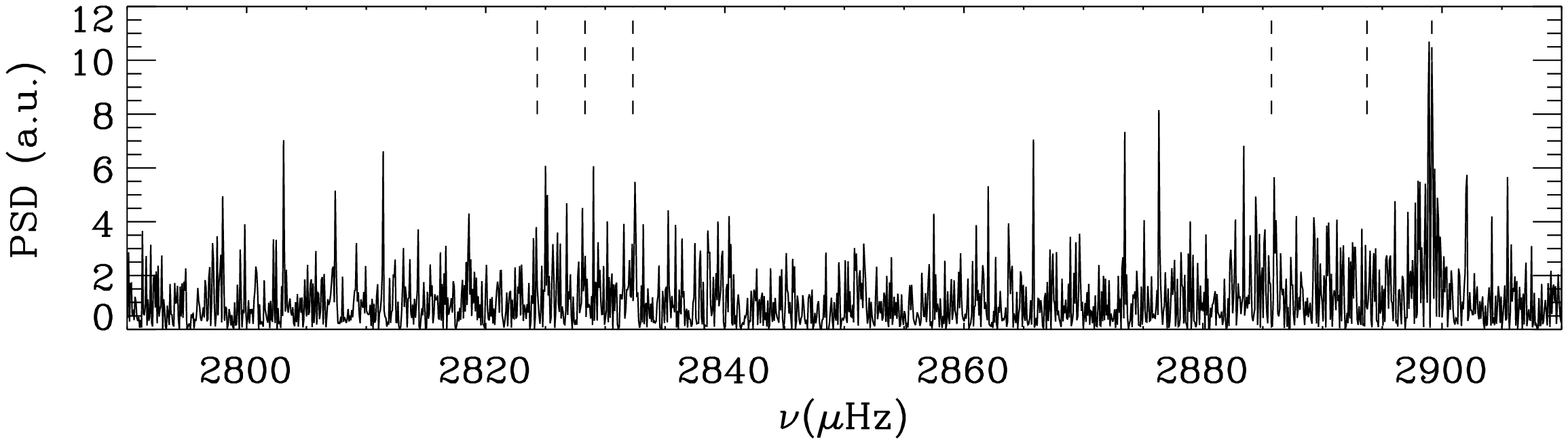}
    	\includegraphics[scale=0.35,angle=90.]{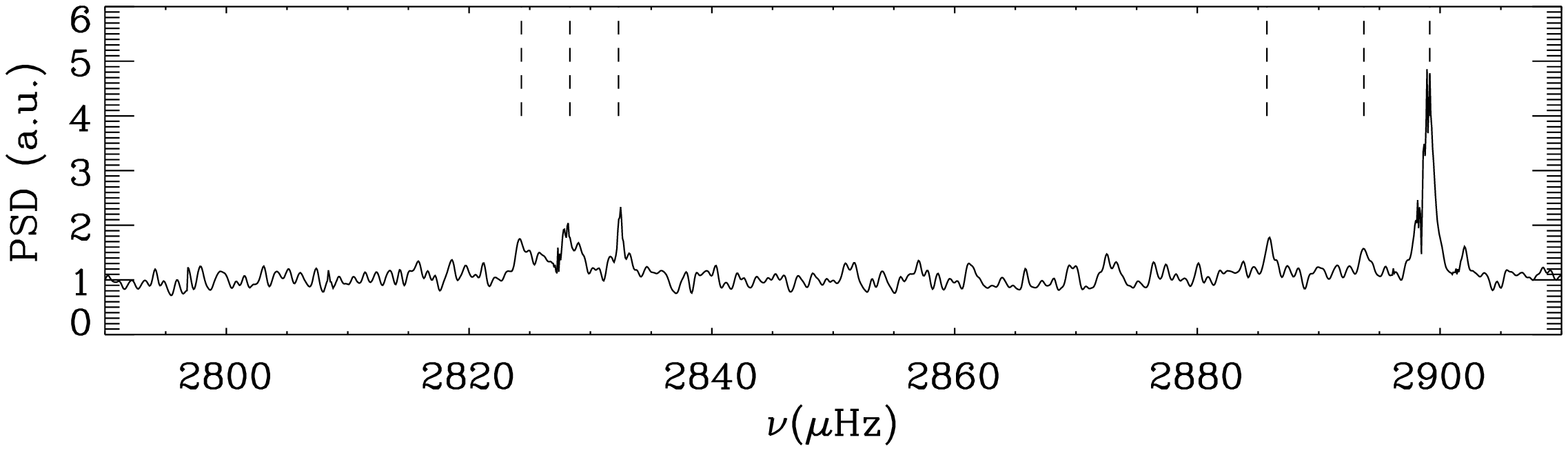}
    	\includegraphics[scale=0.35,angle=90.]{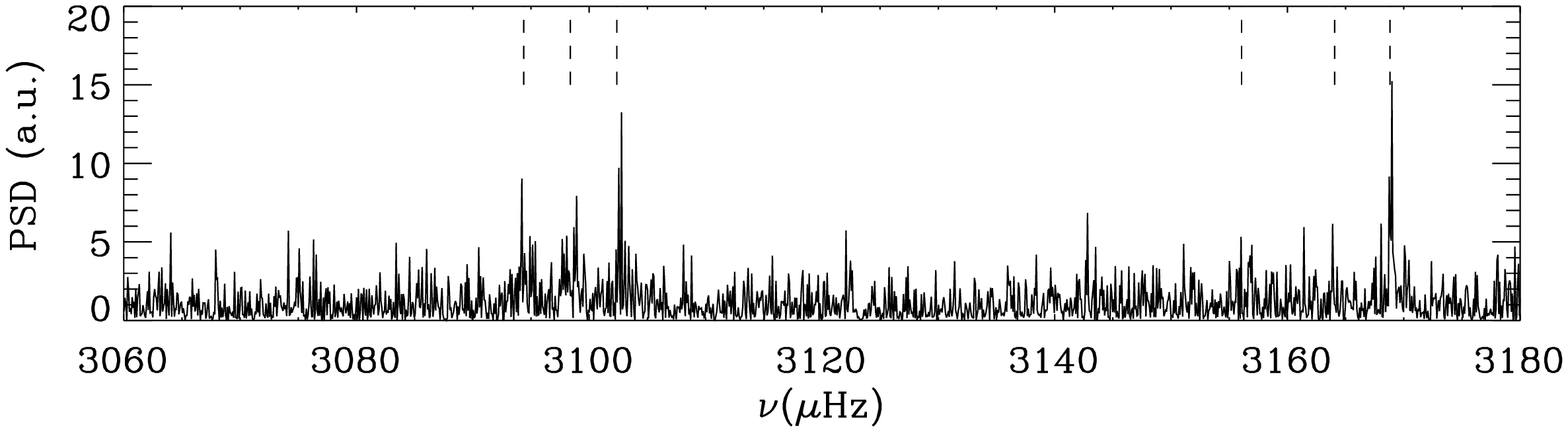}
    	\includegraphics[scale=0.35,angle=90.]{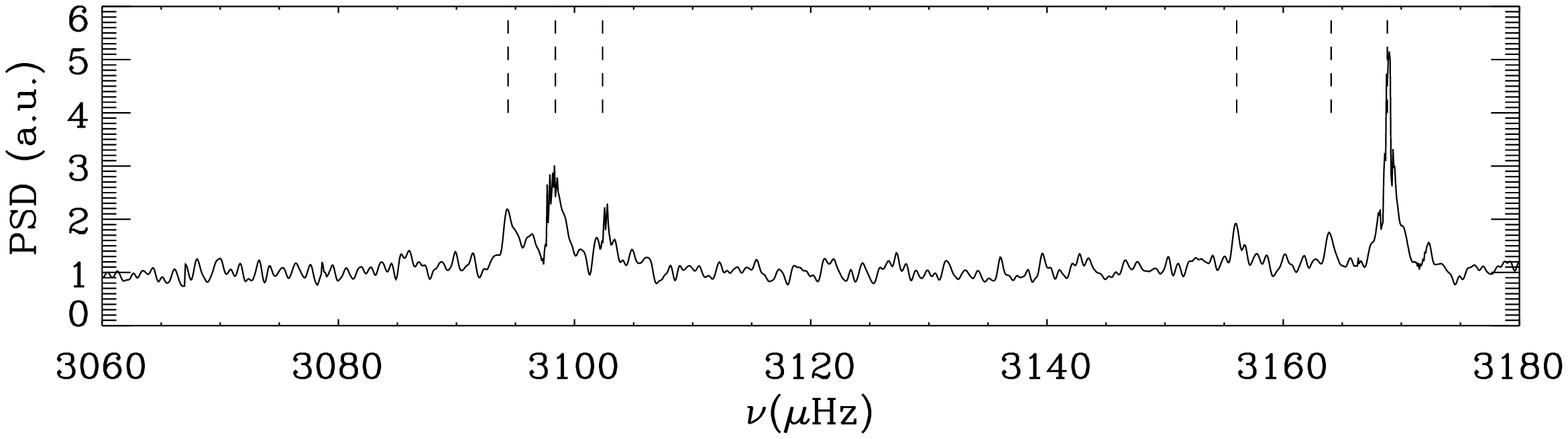}
    	\includegraphics[scale=0.35,angle=90.]{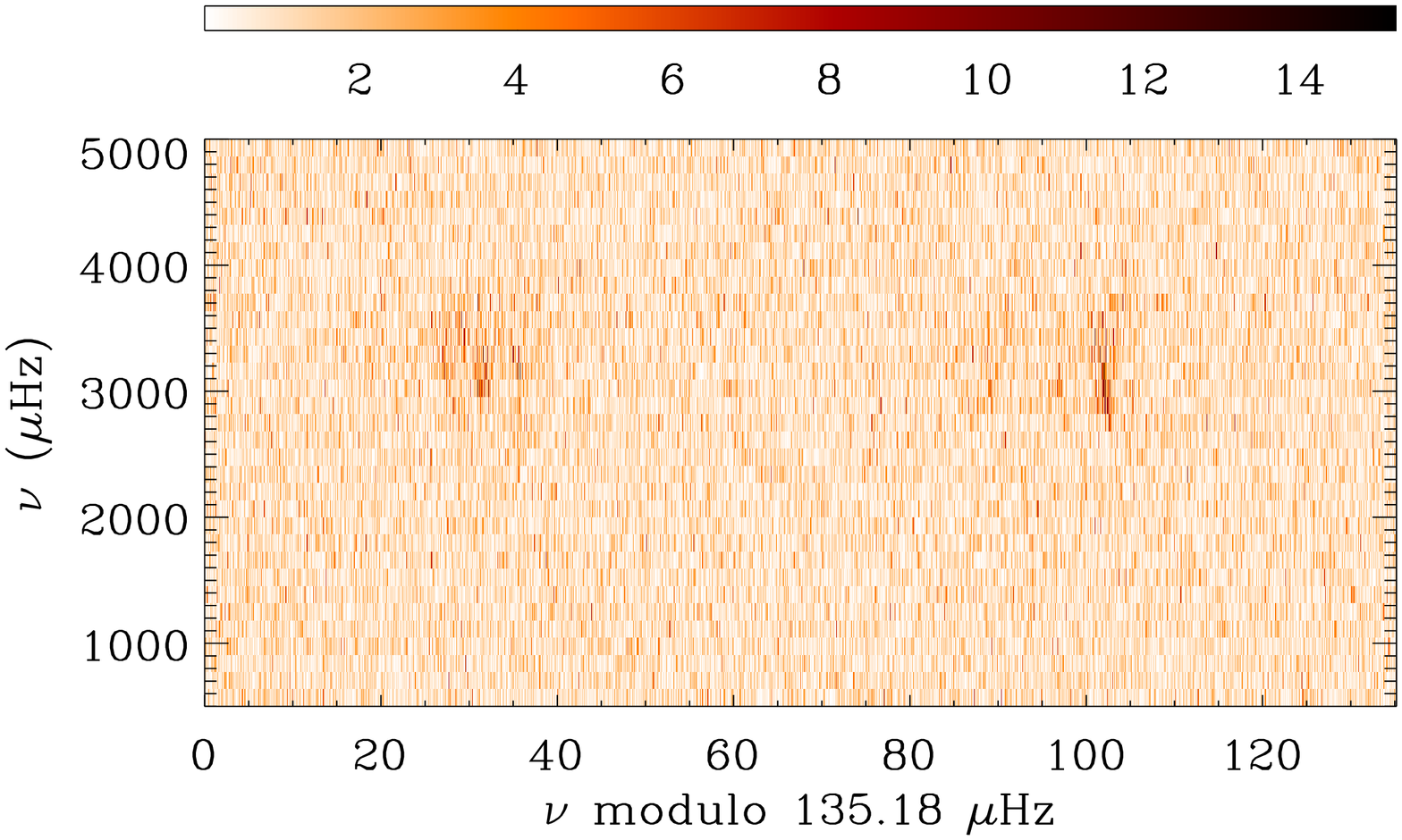}
	\includegraphics[scale=0.35,angle=90.]{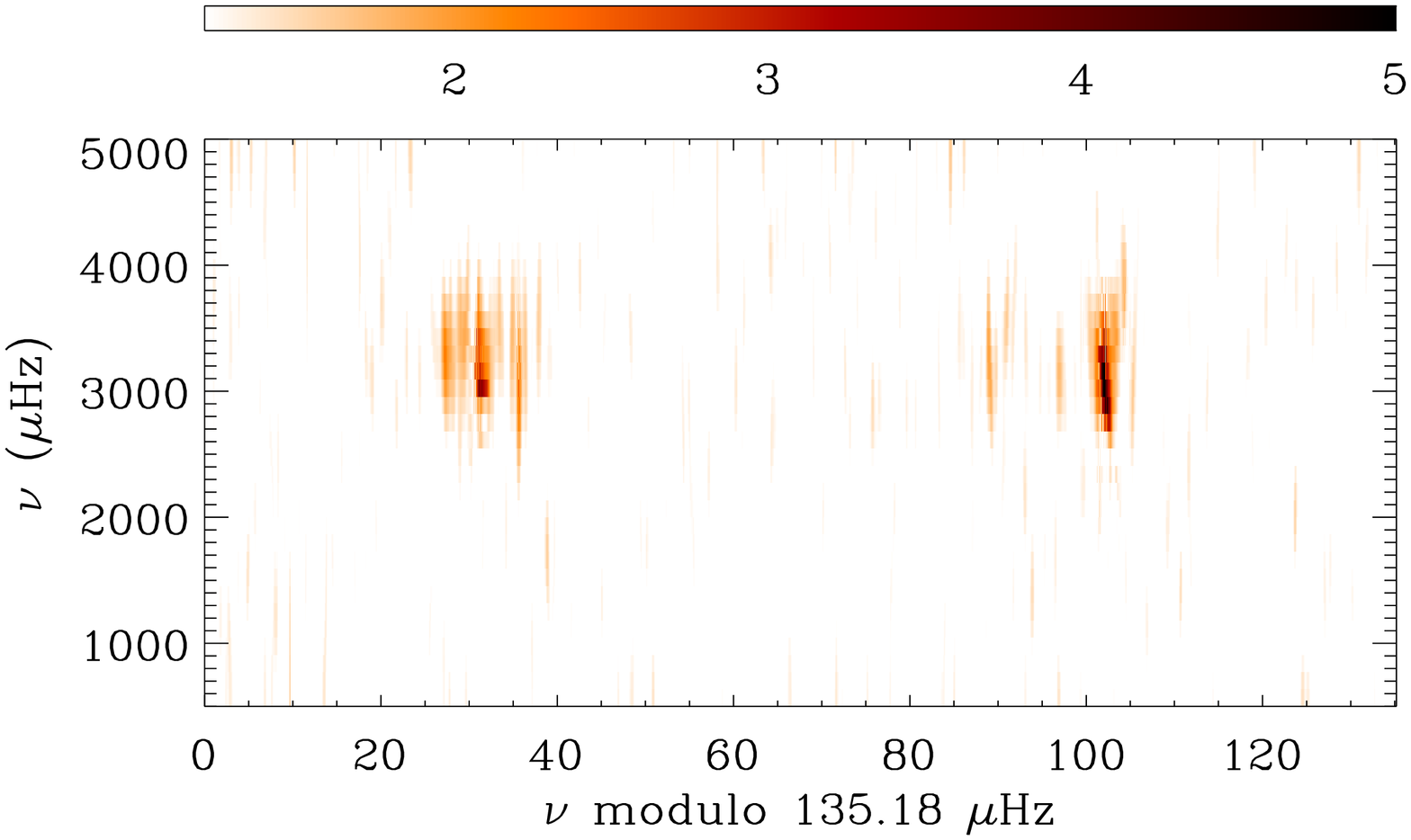}
	\caption{Raw (left) and filtered (right) power spectra (top and middle panels)
and echelle diagrams (bottom panels) for a $S/N=5$ realization. The short dashed
lines in the power spectra represent the position of the theoretical
frequencies. From left to right, the three first equidistant lines indicate the
components $m=-1,0,1$ of $\ell=1$ modes, the two next indicate the strongest
components of $\ell=2$ ($m=-1$ and $1$), and the the last indicates $\ell=0$. In
this case only two components of the $\ell=1$ and the $\ell=0$ mode are slightly
visible in the raw diagram. On the curvelet filtered one, the three $\ell=1$
components appear as well as the $\ell=0$ and the components $m=\pm 1$ of the
$\ell=2$ modes.\label{diagechellesnr5}}
\end{figure*}

Once the spectra have been computed, the echelle diagrams can be built with a
fixed folding frequency. This one corresponds to the mean large frequency
spacing $\Delta\nu_0$, identified either by computing the FFT, the
autocorrelation of the spectra or any other technique \citep[see for
example][]{Regulo2002}. The denoising based on the curvelet transform is then
applied to this echelle diagrams. It is important to note that artifacts may
appear in the filtered spectra at frequencies $\nu^*$=$\nu_0$+$k\Delta\nu_0$,
with $k$ an integer, when random small structures appear in the echelle
diagrams. However, their appearance and position strongly depend on the folding
frequency and are very sensitive to its value. Therefore they can be easily
identified. The artifacts can be reduced (in contrast to the regions containing
signal) by building echelle diagrams with slightly different folding frequencies
and averaging the resultant filtered spectra.

In order to present the results of data analysis using the curvelet denoising
method, we have selected the case of a sun-like star seen with an inclination
angle $i=50\degr$ and with a rotation $\Omega=10 \Omega_{\sun}$. A portion of
the ideal spectra constructed for this star can be seen in Fig.~\ref{theorique}
(top panel). Monte Carlo simulations were then performed, giving rise to
different sets (each one with 500 realizations) of raw spectra with different
$S/N$ ratios. The echelle diagrams were constructed using a folding frequency of
135.18$~\mu$Hz, obtained by computing the FFT of the raw spectrum.

    \subsection{Peak tagging}

In those cases, with a high $S/N$ (typically 15), the mode structure is clearly
visible in each raw spectrum and also on the echelle diagram. The different
ridges can be easily identified and tagged. Although the filtering gives
enhanced denoised diagrams and unfolded spectra, it does not contribute
significantly to the mode identification. 

In the lower $S/N$ cases, however, the situation is different.
Figure~\ref{montecarlo} shows some of the results of the Monte Carlo simulation
for $S/N$=5. 
The upper panel corresponds to 120 realizations among the 500 computed for the
raw spectra in the frequency range 2450--2920$~\mu$Hz. Each horizontal line
corresponds to a single realization. Some patterns can hardly be seen. The lower
panel represents the same spectra after applying the curvelet filtering. A
series of vertical ridges clearly appears. From the left to the right on the
panels, they can be identified as the ($\ell=2$; $m=\pm 1$), the $\ell=0$
(blended with the $\ell=2$; $m=+2$ ) and the ($\ell=1$; $m=-1,0,+1$). The
improvement of the contrast is important in all the realizations and allows to
distinguish the different components of a mode, making easier the identification
and the tagging. 

The identification is harder when looking at each individual spectrum and
requires the use of the echelle diagram. Fig.~\ref{diagechellesnr5} shows an
example of raw (left) and filtered (right) 150-day observation power spectra
(top and middle panels) and the corresponding echelle diagrams (bottom panels)
for a $S/N=5$ realization. Input frequencies are indicated by the short dashed
lines above the spectra. The mode peaks can hardly be distinguished in the raw
spectrum and can easily be confused with noise. For the range 2780-2920$~\mu$Hz,
only a strong peak at 2900$~\mu$Hz can be considered not to be noise. In the
region 3060--3180$~\mu$Hz the peaks are visible and we can attempt to identify
the $\ell$=1 and $\ell$=0 modes but it is still unclear. On the contrary, on the
corresponding parts of the filtered spectrum, the structures of the $\ell$=1
mode with three components, the $\ell$=0 mode and even the strongest components
of the $\ell$=2 mode are visible. 
The raw echelle diagram gives no extra information because of the very weak
ridges and low contrast with the background. The weakest components can hardly
be detected and no tagging can be done. The curvelet filtering provides a
contrast enhancement of the ridges on the echelle diagram. Thus three almost
equidistant strong ridges appear on the left of the diagram and one strong ridge
with two weaker ones on the right. The corresponding patterns can be seen on the
filtered spectrum corresponding well to the theoretical frequencies. Since the
modes $\ell=3$ are not visible, and according to the amplitude of the strongest
peak on the left, we can suggest that the three strongest peaks correspond to a
$\ell=1$ multiplet and the other ones to the $\ell=2$ and $\ell=0$ modes. 

Consequently,  when the tagging is done it is  also easier to have a first
estimation of both the mean rotational splitting and the rotation-axis
inclination, since the visibility of the multiplet is increased. From the
spacing between the components of the mode $\ell=1$, a first estimation of the
mean rotational splitting of the star can be done, as well as an estimation of
the inclination angle, according to their relative amplitude ratios. We have
selected the extraction of one parameter: the mean rotational splitting of the
$\ell$=1 mode at low frequency (2540--2550$~\mu$Hz), to quantify the improvement
of the curvelet filtering. This region is particularly interesting because the
line width is still small and the modes, when they are visible, can be easily
identified. Thus, in a sample of 100 realizations of the Monte Carlo simulation,
we have obtained in 90 of them a better estimation of this parameter in the
filtered spectra. In fact, in the raw spectra it was very exceptional to obtain
a good result. With the filtered spectra a mean rotational splitting of $\langle
\delta\nu \rangle=4.05\pm0.30~\mu$Hz was found, which is very close to the
actual splitting included in the ideal spectra $\langle \delta\nu
\rangle=4.0~\mu$Hz. In addition,  specific methods can be applied to improve the
extraction of these parameters by using different strategies of spectra fitting
as the ones developed by \citet{GizonSolanki2003} or \citet{Ballot2006}. 
In the case of the 30-day observations, the curvelet filtered echelle diagram is
still very noisy and it does not help in recognizing the ridges. However the
corresponding denoised power spectrum is much better despite the lower
resolution (5 times less than in the long runs), even for small $S/N$ ratios
($\sim5$). The modes $\ell=0,2$ and $\ell=1$ can be distinguished, at the
maximum power, while it is not obvious to do so in the raw spectra. Therefore,
we consider that a 
30-day run is the minimum length needed to have reliable results with the
curvelet denoising technique.

\citet{Garcia2005} analyzed the first available MOST public Procyon A data
(32-day observation) using the curvelet technique. Previous analysis by
\citet{Matthews2004} did not reveal the presence of any p-mode structure in this
star. Therefore, due to its tiny S/N ratio the results of the curvelet denoising
should be taken with care. Nevertheless, an excess of power seems to appear in
the region where it is expected and taking the 15 most prominent peaks in this
region, many are in agreement, inside the error bars, with previous tagged modes
using ground-based velocity observations.

\subsection{Extraction of p-mode parameters}

Once the mode identification and tagging are done, the extraction of the mode
parameters can be performed. To illustrate how this extraction can be improved
by using the denoised spectrum we have extracted the central frequency of the
modes in both the raw and the filtered spectra. To determine this parameter,
modes have been fitted by Lorentzian profiles using a maximum-likelihood
estimator in the classical way: adjacent pairs of even ($\ell=0$ and $\ell=2$)
modes are fitted together,  while $\ell=1$ is fitted alone, due to the small
amplitudes of $\ell=3$ modes. For each multiplet, the fitted parameters are the
central frequency $\tilde\nu_{n,\ell}$, the amplitude $\tilde A_{n,\ell}$, the
linewidth $\tilde\Gamma_{n,\ell}$ and the background $b$. The amplitude ratios
inside the multiplets and the rotational splittings have been fixed thanks to
the preliminary estimation done in the previous section (cf. 5.1). The fitting
procedure provides for each adjusted parameter $\tilde{X}$ an associated error
$\sigma(\tilde{X})$ \lastcor{computed by Hessian-matrix inversion}.
 
The raw spectra follow a $\chi^2$ with 2 d.o.f. statistics, whereas the filtered
spectra have a $\chi^2$ with a higher d.o.f. statistics (close to a Gaussian
distribution depending on the number of filtered coefficients). \lastcor{According to \citet{Appourchaux2003}, it is possible to fit spectra following a
$\chi^2$ with more than
2 d.o.f. statistics with a classical procedure developed for a $\chi^2$ with 2
d.o.f. statistics: parameters
are correctly fitted, but computed errors have to be adapted \textit{a
posteriori}. However in our case, 
due to filtering, points of filtred spectra are correlated (we have estimated
that one point is correlated with $\sim$10 neighbouring points). This
correlation should have to be considered, but we have neglected its 
effect on the fitting procedure in the present study. This assumption is
validated by the Monte Carlo simulations.
Such a global filtering induces also correlations between the different lines
of the echelle diagram. Thus the errors on parameters of different modes
(typically $(n,\ell)$ and $(n+1,\ell)$)
can be correlated. These correlations will have to be taken into account
especially during the comparison of frequencies extracted by this way to stellar
models.}

From the 500 realizations of the Monte Carlo simulation, we derived for each
mode and for both the raw and the filtered spectra the mean value of the
extracted frequencies $\langle \tilde\nu_{n,\ell} \rangle$, their mean computed
errors $\langle\sigma(\tilde\nu_{n,\ell})\rangle$ and the dispersion of
frequency distribution $\sigma^{*}(\tilde\nu_{n,\ell})$ (the real error). We
have verified that $\sigma^{*}(\tilde\nu_{n,\ell}) \approx
\langle\sigma(\tilde\nu_{n,\ell})\rangle$ for fits performed on the raw spectra
and \lastcor{we have $\sigma^{*}(\tilde\nu_{n,\ell}) <
\langle\sigma(\tilde\nu_{n,\ell})\rangle$ for fits performed on the filtered
ones. As expected, the error bars on the fitted frequencies, computed by Hessian-matrix inversion, are overestimated.
}

\begin{figure*}
    \centering
    \includegraphics[scale=0.5,angle=90]{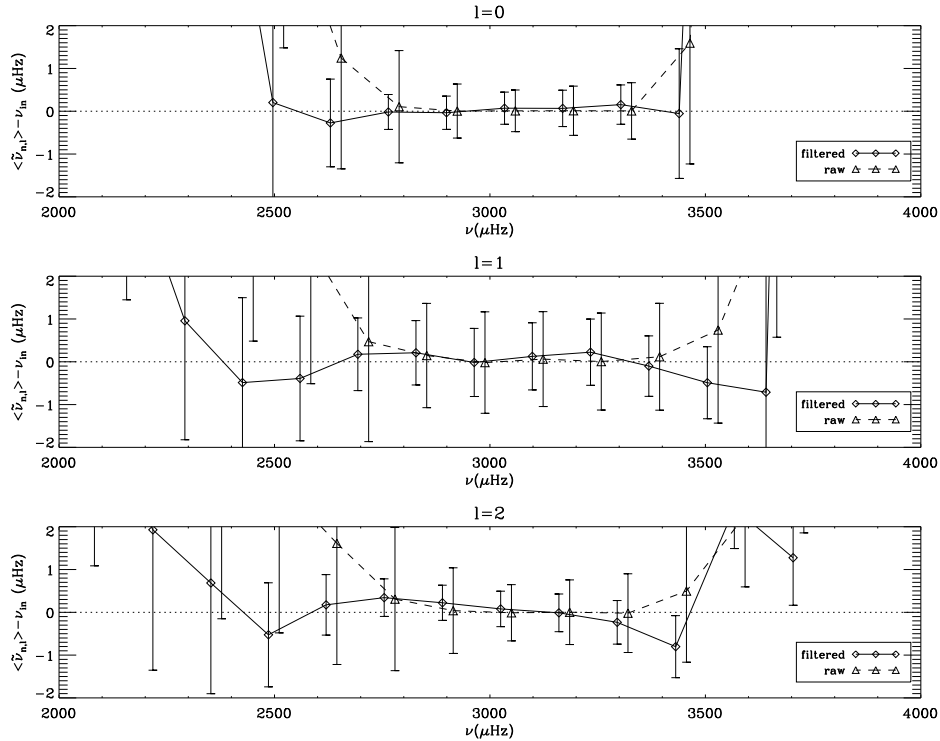}
    \caption{Differences between the mean fitted frequencies $\langle \tilde
\nu_{n,\ell} \rangle$ and the input frequencies $\nu_{in}$, for $\ell=0,1,2$,
for the raw (dashed line with triangles) and filtered (full line with diamonds)
spectra ($S/N=5$, 150-day observation). The error bars correspond to the
dispersion $\sigma^*(\tilde\nu_{n,\ell})$ of the frequency distribution. For
clarity the values for the raw case are shifted by $20~\mu$Hz towards the
right.\label{ecart}}
\end{figure*}

Figure~\ref{ecart} shows the difference between the mean fitted frequencies
$\langle \tilde\nu_{n,\ell} \rangle$ and the theoretical frequencies $\nu_{in}$
of the simulated star discussed in the previous section ($S/N=5$). The error
bars correspond to the dispersion $\sigma^{*}(\tilde\nu_{n,\ell})$. For each
$\ell$, the error bars of the filtered spectra are smaller than those of the raw
spectra. In addition, the range where modes can be detected, tagged and fitted
is extended. While the difference $\langle \tilde\nu_{n,\ell} \rangle -
\nu_{in}$ is only flat in the central region of the raw power spectrum (e.g. for
$\ell=0$, in the range $n=18$--$22$), it extends at higher and lower frequencies
(e.g. for $\ell=0$, the range is extended to $n=16$--$23$) in the filtered one. 

\section{Conclusions}

The application of a noise reduction technique based on the curvelet transform
to echelle diagrams improves the identification -- ``peak tagging'' -- of
stellar acoustic modes. In observations with a $S/N$ ratio as small as 5 we are
still able to recover the mode pattern and extract reliable asteroseismic
information in both small and long runs (30-day and 150-day observations
respectively). Below this S/N and with shorter observations, the method
efficiency is reduced drastically. The rotational splittings and the
rotation-axis inclination can be better estimated using the filtered spectrum.
In particular, Monte Carlo simulations showed that a better extraction of the
mean rotational splitting from modes at low frequency can be done in 90 out of
100 realizations using the filtered spectra. The uncertainty on the extracted
rotational splitting of a typical sun-like star seen with an inclination angle
$i=50\degr$ and with a rotation $\Omega=10 \Omega_{\sun}$ is very small,
$\sim$0.30 $\mu$Hz. These parameters can then be used to have a set of guesses
or \textit{a priori} values to perform individual fits of the spectra. We have
also shown that the range of the frequency extraction can be extended at higher
and lower frequencies using the filtered spectra. Finally, simulations of the
short run observations have demonstrated that this method can also be applied to
lower resolution spectra with good results.

\begin{acknowledgements}
P. Lambert thanks Dr. D. Neuman for useful discussions. 
\end{acknowledgements}
\bibliographystyle{aa}
\bibliography{curvelet}

\begin{thebibliography}{34}
\expandafter\ifx\csname natexlab\endcsname\relax\def\natexlab#1{#1}\fi

\bibitem[{{Anderson} {et~al.}(1990){Anderson}, {Duvall}, \&
  {Jefferies}}]{Anderson1990}
{Anderson}, E.~R., {Duvall}, T.~L., \& {Jefferies}, S.~M. 1990, \apj, 364, 699

\bibitem[{{Appourchaux}(2003)}]{Appourchaux2003}
{Appourchaux}, T. 2003, \aap, 412, 903

\bibitem[{{Baglin} {et~al.}(2001){Baglin}, {Auvergne}, {Catala}, {Michel}, \&
  {COROT Team}}]{Baglin2001}
{Baglin}, A., {Auvergne}, M., {Catala}, C., {Michel}, E., \& {COROT Team}.
  2001, in ESA SP-464: SOHO 10/GONG 2000 Workshop: Helio- and Asteroseismology
  at the Dawn of the Millennium, 395

\bibitem[{{Ballot} {et~al.}(2006){Ballot}, {Garc\'{i}a}, \&
  {Lambert}}]{Ballot2006}
{Ballot}, J., {Garc\'{i}a}, R.~A., \& {Lambert}, P. 2006, MNRAS, {accepted}

\bibitem[{{Ballot} {et~al.}(2004{\natexlab{a}}){Ballot}, {Garc{\'{\i}}a},
  {Lambert}, \& {Teste}}]{BallotYale2004}
{Ballot}, J., {Garc{\'{\i}}a}, R.~A., {Lambert}, P., \& {Teste}, A.
  2004{\natexlab{a}}, in ESA SP-559: SOHO 14 Helio- and Asteroseismology:
  Towards a Golden Future, 309

\bibitem[{{Ballot} {et~al.}(2004{\natexlab{b}}){Ballot}, {Turck-Chi{\`e}ze}, \&
  {Garc{\'{\i}}a}}]{Ballot2004}
{Ballot}, J., {Turck-Chi{\`e}ze}, S., \& {Garc{\'{\i}}a}, R.~A.
  2004{\natexlab{b}}, \aap, 423, 1051

\bibitem[{{Basu} {et~al.}(2004){Basu}, {Mazumdar}, {Antia}, \&
  {Demarque}}]{Basu2004}
{Basu}, S., {Mazumdar}, A., {Antia}, H.~M., \& {Demarque}, P. 2004, \mnras,
  350, 277

\bibitem[{{Bedding} \& {Kjeldsen}(2003)}]{BeddingKjeldsen2003}
{Bedding}, T.~R., \& {Kjeldsen}, H. 2003, Publications of the Astronomical
  Society of Australia, 20, 203

\bibitem[{{Bedding} {et~al.}(2004){Bedding}, {Kjeldsen}, {Butler}, {McCarthy},
  {Marcy}, {O'Toole}, {Tinney}, \& {Wright}}]{Bedding2004}
{Bedding}, T.~R., {Kjeldsen}, H., {Butler}, R.~P., {et~al.} 2004, \apj, 614,
  380

\bibitem[{{Bouchy} \& {Carrier}(2003)}]{BouchyCarrier2003}
{Bouchy}, F., \& {Carrier}, F. 2003, \apss, 284, 21

\bibitem[{{Buzasi} {et~al.}(2000){Buzasi}, {Catanzarite}, {Laher}, {Conrow},
  {Shupe}, {Gautier}, {Kreidl}, \& {Everett}}]{Buzasi2000}
{Buzasi}, D., {Catanzarite}, J., {Laher}, R., {et~al.} 2000, \apjl, 532, L133

\bibitem[{{Cand\`es}(1998)}]{Candes1998}
{Cand\`es}, E.~J. 1998, PhD thesis, {Stanford University}

\bibitem[{{Cand\`{e}s} \& {Donoho}(1999)}]{CandesDonoho1999}
{Cand\`{e}s}, E.~J., \& {Donoho}, D.~L. 1999, in Curves and Surfaces: Saint-Malo
  1999, ed. A. Cohen, C. Rabut, and L. Schumaker (Vanderbilt University Press,
  Nashville, TN)

\bibitem[{{Christensen-Dalsgaard}(2002)}]{JCD2002}
{Christensen-Dalsgaard}, J. 2002, Reviews of Modern Physics, 74, 1073

\bibitem[{{Fierry Fraillon} {et~al.}(1998){Fierry Fraillon}, {Gelly},
  {Schmider}, {Hill}, {Fossat}, \& {Pantel}}]{FierryFraillon1998}
{Fierry Fraillon}, D., {Gelly}, B., {Schmider}, F.~X., {et~al.} 1998, \aap,
  333, 362

\bibitem[{{Garc{\'{\i}}a} {et~al.}(2004){Garc{\'{\i}}a}, {Jim{\'e}nez-Reyes},
  {Turck-Chi{\`e}ze}, {Ballot}, \& {Henney}}]{Garcia2004}
{Garc{\'{\i}}a}, R.~A., {Jim{\'e}nez-Reyes}, S.~J., {Turck-Chi{\`e}ze}, S.,
  {Ballot}, J., \& {Henney}, C.~J. 2004, in ESA SP-559: SOHO 14 Helio- and
  Asteroseismology: Towards a Golden Future, 436

\bibitem[{{Garc\'{i}a} {et~al.}(2006){Garc\'{i}a}, {Lambert}, {Ballot},
  {Pires}, {Nghiem}, {Turck-Chi\`eze}, \& {Matthews}}]{Garcia2005}
{Garc\'{i}a}, R.~A., {Lambert}, P., {Ballot}, J., {et~al.} 2006, A\&A, {submitted}

\bibitem[{{Gizon} \& {Solanki}(2003)}]{GizonSolanki2003}
{Gizon}, L., \& {Solanki}, S.~K. 2003, \apj, 589, 1009

\bibitem[{{Grec} {et~al.}(1983){Grec}, {Fossat}, \& {Pomerantz}}]{Grec1983}
{Grec}, G., {Fossat}, E., \& {Pomerantz}, M.~A. 1983, \solphys, 82, 55

\bibitem[{{Harvey}(1985)}]{Harvey1985}
{Harvey}, J. 1985, in Future Missions in Solar, Heliospheric and Space Plasma
  Physics, 199

\bibitem[{{Ledoux}(1951)}]{Ledoux1951}
{Ledoux}, P. 1951, \apj, 114, 373

\bibitem[{{Lopes} \& {Turck-Chi\`{e}ze}(1994)}]{Lopes1994}
{Lopes}, I., \& {Turck-Chi\`{e}ze}, S. 1994, \aap, 290, 845

\bibitem[{{Matthews}(1998)}]{Matthews1998}
{Matthews}, J.~M. 1998, in Structure and Dynamics of the Interior of the Sun
  and Sun-like Stars SOHO 6/GONG 98 Workshop Abstract, June 1-4, 1998, Boston,
  Massachusetts, 395

\bibitem[{{Matthews} {et~al.}(2004){Matthews}, {Kusching}, {Guenther},
  {Walker}, {Moffat}, {Rucinski}, {Sasselov}, \& {Weiss}}]{Matthews2004}
{Matthews}, J.~M., {Kusching}, R., {Guenther}, D.~B., {et~al.} 2004, \nat, 430,
  51

\bibitem[{{Monteiro} {et~al.}(2000){Monteiro}, {Christensen-Dalsgaard}, \&
  {Thompson}}]{Monteiro2000}
{Monteiro}, M.~J.~P.~F.~G., {Christensen-Dalsgaard}, J., \& {Thompson}, M.~J.
  2000, \mnras, 316, 165

\bibitem[{{R{\'e}gulo} \& {Roca Cort{\'e}s}(2002)}]{Regulo2002}
{R{\'e}gulo}, C., \& {Roca Cort{\'e}s}, T. 2002, \aap, 396, 745

\bibitem[{{Roxburgh} \& {Vorontsov}(2000{\natexlab{a}})}]{Roxburgh2000a}
{Roxburgh}, I.~W., \& {Vorontsov}, S.~V. 2000{\natexlab{a}}, \mnras, 317, 141

\bibitem[{{Roxburgh} \& {Vorontsov}(2000{\natexlab{b}})}]{Roxburgh2000b}
{Roxburgh}, I.~W., \& {Vorontsov}, S.~V. 2000{\natexlab{b}}, \mnras, 317, 151

\bibitem[{{Starck} {et~al.}(2004){Starck}, {Aghanim}, \& {Forni}}]{Starck2004}
{Starck}, J.~L., {Aghanim}, N., \& {Forni}, O. 2004, \aap, 416, 9

\bibitem[{{Starck} {et~al.}(2002){Starck}, {Cand\`es}, \&
  {Donoho}}]{Starck2002}
{Starck}, J.~L., {Cand\`es}, E.~J., \& {Donoho}, D.~L. 2002, IEEE Transactions
  on Image Processing, 11, 670

\bibitem[{{Starck} {et~al.}(2003){Starck}, {Donoho}, \&
  {Cand{\`e}s}}]{Starck2003}
{Starck}, J.~L., {Donoho}, D.~L., \& {Cand{\`e}s}, E.~J. 2003, \aap, 398, 785

\bibitem[{{Tassoul}(1980)}]{Tassoul1980}
{Tassoul}, M. 1980, \apjs, 43, 469

\bibitem[{{Turck-Chi{\`e}ze} {et~al.}(1993){Turck-Chi{\`e}ze}, {D{\"a}ppen},
  {Fossat}, {Provost}, {Schatzman}, \& {Vignaud}}]{TurckChieze1993}
{Turck-Chi{\`e}ze}, S., {D{\"a}ppen}, W., {Fossat}, E., {et~al.} 1993,
  \physrep, 230, 57

\bibitem[{{Vorontsov}(1991)}]{Vorontsov1991}
{Vorontsov}, S.~V. 1991, Soviet Astronomy, 35, 400

\end{thebibliography}
\end{document}